\documentclass{article}
\usepackage{frascatiphys,here,graphicx,subfigure}
\begin{document}
\title{ 
\bf CHARM PHYSICS: HINTS FOR A MATURE DESCRIPTION OF HADRONS
}
\author{
A. Valcarce \\
{\em Departamento de F\'\i sica Fundamental, Universidad de Salamanca, Spain} \\
J. Vijande \\
{\em Departamento de F\' \i sica Te\'orica,
Universidad de Valencia, Spain}\\
N. Barnea \\
{\em The Racah Institute of Physics, The Hebrew University,
Jerusalem, Israel}
}
\maketitle
\baselineskip=11.6pt
\begin{abstract}
The physics of charm has become one of the best laboratories exposing the
limitations of the naive constituent quark model and also giving hints
into a more mature description of hadron spectroscopy. Recent discoveries
are a challenge that have revolutionized our understanding of the hadron 
spectra. In this talk we address the study of many-quark components in 
charmonium spectra. To make the physics clear we also discuss 
exotic many-quark systems.
\vspace*{0.3cm}
\end{abstract}
\baselineskip=14pt
More than thirty years after the so-called November revolution\cite{Bjo85}, 
heavy meson spectroscopy is being again a challenge. The formerly comfortable 
world of heavy meson espectroscopy is being severely tested by new 
experiments\cite{Ros07}. This challenging situation arose in 
the open-charm sector
with the discovery of the $D_{sJ}^*(2317)$, the $D_{sJ}(2460)$ and the
$D_0^*(2308)$ mesons. All of them are positive parity states with masses smaller
than expectations from quark potential models, and in the first two cases
also smaller widths. In general, one could say that 
the area phenomenologically understood in the open-charm
meson spectrum extends to
states where the $q\bar q$ pair is in relative $S-$wave. In the positive
parity sector, $P-$wave states, is where the problems arise. This has been said
as an example where naive quark models are probably too naive\cite{Clo07}. 
Out of the many explanations suggested for these states, 
the unquenching of the naive quark model has been successful\cite{Vij05}. 
When a $q\bar q$ pair occurs in a $P-$wave but can couple to hadron 
pairs in $S-$wave the latter will distort the $q\bar q$ picture. 
In the examples mentioned above, the $0^+$ and $1^+$ $c\bar s$ states
predicted above the $DK(D^*K)$ thresholds couple to the continuum. This 
mixes $DK(D^*K)$ components in the wave function.
This idea can be easily formulated in terms of a meson wave-function
described by 
\begin{equation}
\label{mes-w}
\left|\psi\right>=\sum_{i} \alpha_i \left|q
\bar q\right>_i + \sum_{j} \beta_j \left|qq\bar q \bar q\right>_j
\end{equation}
where $q$ stands for quark degrees of
freedom and the coefficients $\alpha_i$ and $\beta_j$ take into
account the possible admixture of four-quark components in the
standard $q \bar q$ picture.

This explanation has open the discussion about the presence of compact 
four-quark states in charmonium spectroscopy. This is an old 
idea long ago advocated to explain the proliferation
of light-scalar mesons\cite{Jaf07}. In the case of charmonium
spectroscopy, some members of the new hadronic zoo may fit in the 
simple quark model description as $q\bar q$  pairs 
($X(3940)$, $Y(3940)$, and $Z(3940)$ may fit
into the $\chi_{c0}$, $\chi_{c1}$, and $\chi_{c2}$ quark model structure)
others appear to be more elusive ($X(3872)$ and $Y(4260)$). 

The debate has been open with special emphasis on the nature of the $X(3872)$.
Since it was first reported by Belle in 2003\cite{Bel03} 
it has gradually become the flagship of a new armada of states 
whose properties make their identification as
traditional $q\bar q$ states unlikely. In this 
heterogeneous group we could include states like
the $Y(2460)$ reported by $BABAR$,
and the aforementioned $D_{sJ}(2317)$ and $D_{sJ}(2460)$ 
reported by $BABAR$ and CLEO. 
An average mass of 3871.2$\pm$0.5 MeV and a narrow width of less than 2.3 MeV
have been reported for the $X(3872)$.
Note the vicinity of this state to the $D^0\overline{D^{*0}}$ threshold,
$M(D^0\,\overline{D^{*0}})=3871.2\pm1.2$ MeV.
With respect to the $X(3872)$ quantum numbers, neither D0 
nor $BABAR$ have been able to offer a clear prediction.
Its isovector nature has been 
excluded by $BABAR$ due to the negative results
in the search for a charged partner in the decay 
$B\to X(3872)^-K$, $X(3872)^-\to J/\psi\pi^-\pi^0$\cite{Bab05c}. 
CDF has studied the $X(3872)$ $J^{PC}$ quantum numbers using
dipion invariant mass distribution and angular analysis, 
obtaining that only the assignments $1^{++}$ and
$2^{-+}$ are able to describe data\cite{CDF06}.
On the other hand, recent studies by Belle 
combining angular and kinematic properties
of the $\pi^+\pi^-$ invariant mass
strongly favor a $J^{PC}=1^{++}$ state,
and the observation of the $X(3872)\to D^0\overline{D^0}\pi^0$ also
prefers the $1^{++}$ assignment compared to the $2^{-+}$\cite{Bel06}.
Therefore, although some caution is still required until better statistic
is obtained\cite{Set06}, an isoscalar $J^{PC}=1^{++}$ state seems to
be the best candidate to describe the properties of the $X(3872)$.
\begin{table}[t]
\centering
\caption{ \it $c\bar c n\bar n$ results. 
}
\vskip 0.1 in
\begin{tabular}{|c|cc|cc|} \hline
 &\multicolumn{2}{|c|}{CQC} &\multicolumn{2}{|c|}{BCN} \\
\hline
$J^{PC}(K_{\rm max})$ & $E_{4q}$ & $\Delta_{E}$&
$E_{4q}$ & $\Delta_{E}$ \\ 
\hline
\hline
$0^{++}$ (24) & 3779 &  +34 &  3249 &  +75  \\
$0^{+-}$ (22) & 4224 &  +64 &  3778 & +140  \\
$1^{++}$ (20) & 3786 &  +41 &  3808 & +153  \\
$1^{+-}$ (22) & 3728 &  +45 &  3319 &  +86  \\
$2^{++}$ (26) & 3774 &  +29 &  3897 &  +23  \\
$2^{+-}$ (28) & 4214 &  +54 &  4328 &  +32  \\
$1^{-+}$ (19) & 3829 &  +84 &  3331 & +157  \\
$1^{--}$ (19) & 3969 &  +97 &  3732 &  +94  \\
$0^{-+}$ (17) & 3839 &  +94 &  3760 & +105  \\
$0^{--}$ (17) & 3791 & +108 &  3405 & +172  \\
$2^{-+}$ (21) & 3820 &  +75 &  3929 &  +55  \\
$2^{--}$ (21) & 4054 &  +52 &  4092 &  +52 \\
\hline
\end{tabular}
\label{t1}
\end{table}

\begin{figure}[t]
\vspace*{-5.5cm}
    \begin{center}
        {\includegraphics[width=\textwidth]{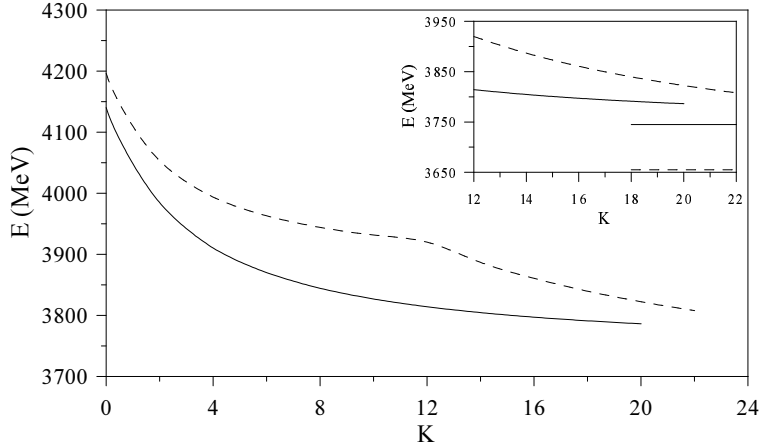}}
\vspace*{-6cm}
\caption{\it Energy of the $1^{++}$ state using the CQC (solid line)
and BCN models (dashed line) as a function of $K$. The insert
in the upper-right corner magnifies the large values of $K$
to show the convergence to the corresponding threshold showed
by a straight line.}
\label{f1}
    \end{center}
\end{figure}

\begin{figure}[t]
\vspace*{-1cm}
    \begin{center}
        {\includegraphics[width=8cm,height=8cm]{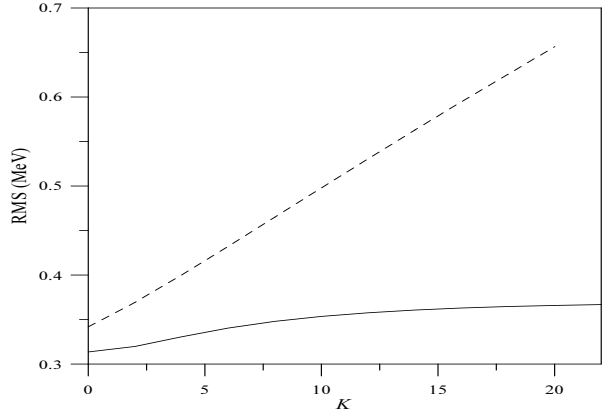}}
\vspace*{-1cm}
\caption{\it Evolution with $K$ of the radius (RMS) of the $c\bar c n\bar n$
  $J^{PC}=1^{++}$
state (dashed line) and the $cc\bar n \bar n$ $J^P=1^+$
state (solid line) for the CQC model.}
\label{f2}
    \end{center}
\end{figure}

To study the possible existence of four-quark states in the
charmonium spectrum 
we have solved exactly the four-body Schr\"odinger
equation using the hyperspherical harmonic (HH) formalism\cite{Vij07}.
We have used two standard quark-quark interaction models: a potential
containing a linear confinement and a Fermi-Breit one-gluon exchange
interaction (BCN), and a potential containing besides boson exchanges
between the light quarks (CQC). The model parameters have been tuned
in the meson and baryon spectra. To make the physics clear we have 
solved simultaneously two different type of systems: the cryptoexotic
$c\bar c n \bar n$ and the flavor exotic $c c \bar n \bar n$, where
$n$ stands for a light $u$ or $d$ quark. The results are reported in
Tables~\ref{t1} and \ref{t2}, indicating the quantum numbers of the
state studied, $J^{PC}$,
the maximum value of the grand angular momentum used in the HH 
expansion, $K_{max}$, and the
energy difference between the mass of the 
four-quark state, $E_{4q}$, and that of the lowest two-meson
threshold calculated with the same potential model, $\Delta_E$. For
the $cc\bar n \bar n$ system we have also calculated 
the radius of the four-quark
state, $R_{4q}$, and its ratio to the sum of the radii of the 
lowest two-meson threshold, $R_{4q}/(r^1_ {2q}+r^2_{2q})$. 
\begin{table}[t]
\centering
\caption{ \it $cc\bar n\bar n$ results. 
}
\vskip 0.1 in
\begin{tabular}{|c|c|cccc|} \hline
 & &\multicolumn{4}{|c|}{CQC} \\
\hline
& $J^{P}(K_{\rm max})$ & $E_{4q}$ & $\Delta_{E}$ &
$R_{4q}$ & $R_{4q}/(r^1_ {2q}+r^2_{2q})$ \\
\hline
\hline
 & $0^{+}$ (28) & 4441 &  +15 &  0.624 & $> 1$ \\
 & $1^{+}$ (24) & 3861 &$-$76 &  0.367 & 0.808 \\
I=0 & $2^{+}$ (30) & 4526 &  +27 &  0.987 & $> 1$ \\
 & $0^{-}$ (21) & 3996 &  +59 &  0.739 & $> 1$ \\
 & $1^{-}$ (21) & 3938 &  +66 &  0.726 & $> 1$ \\
 & $2^{-}$ (21) & 4052 &  +50 &  0.817 & $> 1$ \\
\hline
 & $0^{+}$ (28) & 3905 &  +50 &  0.817 & $> 1$ \\
 & $1^{+}$ (24) & 3972 &  +33 &  0.752 & $> 1$  \\
I=1 & $2^{+}$ (30) & 4025 &  +22 &  0.879 & $> 1$ \\
 & $0^{-}$ (21) & 4004 &  +67 &  0.814 & $> 1$ \\
 & $1^{-}$ (21) & 4427 &  +1  &  0.516 & 0.876 \\
 & $2^{-}$ (21) & 4461 &$-$38 &  0.465 & 0.766 \\
\hline
\end{tabular}
\label{t2}
\end{table}
As can be seen in Table~\ref{t1}, in the case of the $c\bar c n\bar n$
there appear no bound states for any set of quantum numbers, including
the suggested assignments of the $X(3872)$: $1^{++}$ and $2^{-+}$. 
The situation is different for the
$cc\bar n \bar n$ where we observe the existence of bound states.
It is particularly interesting the $J^P=1^+$ channel, that it is bound
both with the CQC and the BCN models.
For the $c\bar c n\bar n$ system, independently
of the quark-quark interaction and the quantum numbers 
considered, the system evolves to a
well separated two-meson state. This is clearly seen
in the energy, approaching the corresponding two free-meson threshold,
and also in the probabilities of the
different color components of the wave function
and in the radius. 
We illustrate the convergence plotting 
in Fig.~\ref{f1} the energy of the $J^{PC}=1^{++}$ state 
as a function of $K$. It can be observed how 
the BCN $1^{++}$ state does not converge to the lowest
threshold for small values of $K$, being affected by the 
presence of an intermediate $J/\psi\,\omega\vert_S$
threshold with an energy of 3874 MeV. Once sufficiently 
large values of $K$ are considered the system follows
the usual convergence to the lowest threshold 
(see insert in Fig.~\ref{f1}).
The dashed line of Fig.~\ref{f2} illustrates how the system evolves to
two singlet color mesons, whose separation increases with $K$.
Thus, in any manner one can claim for the existence
of a bound state for the $c\bar c n \bar n$ system. 

A completely different behavior is observed in Table~\ref{t2}.
Here, there are some particular quantum numbers 
where the energy is quickly stabilized below
the theoretical threshold.
For example, the solid line in Fig.~\ref{f2} illustrates how
the radius of the $1^+$ $cc\bar n\bar n$ state
is stable, and it is smaller than the
sum of the radius of the two-meson threshold. 
We obtain $r_{4q}=0.37$ fm
compared to $r_{M_1}+r_{M_2}= 0.44$ fm for the $1^+$ state. 
The analysis of the color components in the wave function
is involved in this case. 
One cannot directly conclude
the presence of octet-octet components in the wave function,
because the octet-octet color component in the
$(c_1\bar n_3)(c_2\bar n_4)$ basis can be re-expressed
as a singlet-singlet color component in the
$(c_1\bar n_4)(c_2\bar n_3)$ coupling, being the
same physical system due to the identity of the two
quarks and the two antiquarks. 
The actual interest
and the capability of some experiments\cite{Sel05}
to detect double charmed states makes this prediction
a primary objective to help in the understanding of QCD
dynamics.

There is an important difference between the 
two physical systems studied.
While for the $c\bar c n\bar n$ there are 
two allowed physical {\it decay channels},
$(c\bar c)(n\bar n)$ and $(c\bar n)(\bar c n)$, for the $cc\bar n\bar n$
only one physical system contains 
the possible final states, $(c \bar n)(c\bar n)$.
This has important consequences if both systems (two- and four-quark
states) are described within the same two-body Hamiltonian,
the $c \bar c n \bar n$ will hardly present bound states, because the system
will reorder itself to become the lightest two-meson state, either
$(c\bar c)(n\bar n)$ or $(c\bar n)(\bar c n)$. In other words,
if the attraction is provided by the interaction between
particles $i$ and $j$, it does also contribute to the asymptotic
two-meson state. This does not happen
for the $c c\bar n\bar n$ if the interaction between, for example,
the two quarks is strongly attractive. 
In this case there is no asymptotic two-meson
state including such attraction, and therefore the system will bind.

Once all possible quantum numbers of the $X(3872)$ have 
been analyzed and discarded very
few alternatives remain. If this state is 
experimentally proved to be a compact four-quark state
this will point either to the existence of non two-body 
forces or to the emergence of
strongly bound diquark structures within the tetraquark. 
Both possibilities are appealing, does the interaction
becomes more involved with the number of quark or does 
the Hilbert space becomes simpler? On the one hand, some
lattice QCD collaborations\cite{Oki05} have reported 
the important role played by three- and four-quark
interactions within the confinement (the $Y-$ and $H-$shape). 
On the other hand, diquark correlations
have been proposed to play a relevant role in several 
aspects of QCD, from baryon spectroscopy to
scaling violation\cite{Jaf03}. The spontaneous formation 
of diquark components can be checked within our formalism.
The four-quark state can be explicitly written in the $(cn)(\bar c\bar n)$
coupling to isolate the diquark-antidiquark configurations. In the
case of $J^{PC}=1^{++}$ 
only two components of the wave function have the proper quantum numbers 
to be identified with a diquark, being their total probability
less than 3\%. Therefore, it is clear that 
without any further hypothesis two-body
potentials do not favor the presence of diquarks and any
description of these states in terms of diquark-antidiquark
components would be selecting a restricted Hilbert space.

Finally, our conclusions can be made more general. If we have an $N$-quark
system described by two-body interactions in such a way that there exists
a subset of quarks that cannot make up a physical subsystem, then one may expect
the existence of $N$-quark bound states by means of central two-body potentials.
If this is not true one will hardly find $N-$quark bound states\cite{Lip75}.
For the particular case of the tetraquarks,
this conclusion is exact if the confinement is described
by the first $SU(3)$ Casimir operator, because when the system is split
into two-mesons the confining contribution from the two isolated mesons
is the same as in the four-quark system.
The contribution of three-body color forces\cite{Dmi05}
would interfere in the simple
comparison of the asymptotic and the compact states.
Another possibility in the same line would be a modification
of the Hilbert space. If for some reason particular components of the
four-quark system (diquarks) would be favored against others, the
system could be compact\cite{Mai04}. Lattice QCD calculations\cite{Ale07}
confirm the phenomenological expectation that QCD dynamics favors the
formation of good diquarks\cite{Jaf07}, i.e., in the scalar positive
parity channel. However, they are large objects whose relevance to hadron
structure is still under study. All these alternatives will allow
to manage the four-quark system without affecting the threshold and
thus they may allow to generate any solution.

%
This work has been partially funded by MCyT
under Contract No. FPA2007-65748 and by JCyL
under Contract No. SA016A07.
%

%
\end{document}